\documentclass[reprint,
amsmath,amssymb, superscriptaddress
]{revtex4-1}

\usepackage[utf8]{inputenc}
\usepackage[english]{babel}
\usepackage{float}
\usepackage{xcolor}
\usepackage{graphicx}
\usepackage{dcolumn}
\usepackage{bm}
\usepackage[hidelinks]{hyperref}

\usepackage{pdfpages}

\makeatletter
\AtBeginDocument{\let\LS@rot\@undefined}
\makeatother

\hypersetup{
    colorlinks=true,
    linkcolor=blue,
    citecolor=blue,
    filecolor=blue,
    urlcolor=blue,
}

\begin{document}

%\preprint{APS/123-QED}
\preprint{AIP/123-QED}

\title{Anisotropic second-harmonic generation in superconducting nanostructures}

 \author{Sara Memarzadeh}
 \email{sara.memarzadeh@amu.edu.pl}
 \affiliation{Institute of Spintronics and Quantum Information, \\Faculty of Physics and Astronomy, Adam Mickiewicz University Poznań, \\Uniwersytetu Poznańskiego 2, 61-614 Poznań, Poland}
 
 \author{Maciej Krawczyk}
 \affiliation{Institute of Spintronics and Quantum Information, \\Faculty of Physics and Astronomy, Adam Mickiewicz University Poznań, \\Uniwersytetu Poznańskiego 2, 61-614 Poznań, Poland}

 \author{Armen Gulian}
 \affiliation{Laboratory of Advanced Quantum Materials and Devices, Institute for Quantum Studies, Chapman University, \\One University Drive, Orange, California, 92866, United States of America}

 \author{Jarosław W. Kłos}
 \affiliation{Institute of Spintronics and Quantum Information, \\Faculty of Physics and Astronomy, Adam Mickiewicz University Poznań, \\Uniwersytetu Poznańskiego 2, 61-614 Poznań, Poland}

\begin{abstract}{
Circuits based on superconducting nanostructures are among the most promising platforms for quantum computing. Understanding how device geometry governs nonlinear electrodynamics is crucial for implementing superconducting quantum technologies. However, to date, research has largely been limited to superconducting nanostructures with collinearly aligned static and dynamic applied magnetic fields. Here, we analyze the dynamics of Meissner currents and Abrikosov vortices in a superconducting nanocube exposed to combined static and microwave magnetic fields, extending the analysis to a more general excitation geometry. We demonstrate that, in a noncollinear configuration, the magnetization component parallel to the static field develops a dominant second-harmonic response under the microwave driving. This effect is strongly enhanced when Meissner currents saturate at static fields just below the thresholds for successive vortex nucleation. By numerically solving the time-dependent Ginzburg–Landau equations, we show that the response originates from Meissner-current saturation combined with the nonlinear oscillations of normal-phase indentations, yielding an anisotropic second-harmonic signal that is directionally separated from, and not overshadowed by, the first-harmonic component of the dynamic magnetization. These findings are relevant for superconducting devices that require controllable high-frequency nonlinearity.
}\end{abstract}

%\keywords{Suggested keywords} 

\maketitle

\section{Introduction \label{sec:Intro}}

Superconductors operate at cryogenic temperatures where thermal fluctuations are minimal~\cite{Tinkham1996}, providing a clean electromagnetic environment that enables clearer isolation of nonlinear effects. Furthermore, the dynamics of supercurrents can be precisely controlled by external fields and geometry~\cite{D4NH00618F}, allowing systematic investigation of second-harmonic generation (SHG) mechanisms in well-defined settings. 
Indeed, SHG has been both theoretically predicted and experimentally observed under various conditions, including vortex motion~\cite{Eley_2021} in response to oscillatory fields~\cite{PhysRevLett.133.036004}, inversion symmetry breaking in unconventional superconductors~\cite{PhysRevB.100.220501}, and mixed-parity pairing in non-centrosymmetric systems with strong spin–orbit coupling~\cite{PhysRevB.102.184516, Miclea2012}. These findings show 
that SHG can act as a sensitive probe for detecting subtle symmetry-breaking phenomena and vortex-related dynamics.

Nonlinear superconducting responses are technologically relevant in fluxonic devices~\cite{dobrovolskiy_2020, baumgartner2022}, where broken symmetries enable nonreciprocal transport, vortex-ratchet effects, and superconducting-diode behavior~\cite{Wakatsuki_2017, Porrati_2025}. In superconducting sensing and computing~\cite{Clarke2008}, however, the presence of vortices can limit the lifetime of the quantum state by introducing dissipation and noise~\cite{olivers_2013}, and SHG can serve as a signature of these processes.

Nonlinear effects in superconductors, including the generation of higher harmonics in their electromagnetic response, were reported in the 1960s~\cite{Bean_1964}. They are closely linked to the hysteretic magnetization response, which, according to the critical-state model, originates from the critical current density and magnetic flux trapping during field reversal~\cite{Sohn_1989,Chaddah_1992, Shantsev_2000}. 
The critical-state model predicts that a second harmonic in the microwave response can arise when inversion symmetry is broken by an applied static magnetic field~\cite{Sohn_1989}, consistent with the general requirement of symmetry breaking for SHG~\cite{Franken_1963,Boyd_2008}.
In this model~\cite{Bean_1964}, nonlinearity originates from finite Meissner screening currents. However, in type-II superconductors, vortex dynamics may also contribute to nonlinearity and SHG. Inversion-symmetry breaking can result from the introduction of transport currents \cite{Nakamura_2020} or the presence of screening currents~\cite{Nakamura_2024}, which modify the vortex pinning potential and induce anharmonic vortex motion under a driving field. Such effects are particularly relevant in systems where the effective vortex mass is small and the inertial term in the oscillator model outweighs viscous damping, allowing fast THz responses with SHG~\cite{Nakamura_2024}.

The nonlinear alternating current (AC) magnetic response of type-II superconductors has long served as a powerful probe of vortex dynamics under oscillatory magnetic fields. 
Early theoretical developments established a macroscopic framework that unified the linear and nonlinear regimes of the AC response, revealing that higher harmonics arise as clear indicators of vortex depinning when the driving frequency remains below the characteristic pinning 
frequency~\cite{PhysRevB.48.3393}. 
Subsequent experimental studies on Hg-based high-$T_c$ materials confirmed this picture under parallel direct current (DC) and AC magnetic fields, where the emergence of second and third harmonics was related to irreversible vortex motion arising from bulk pinning and surface barriers~\cite{PhysRevB.61.9793}. 
More recently, low-frequency mutual inductance measurements on MoGe and NbN thin films have revealed pronounced nonlinear magnetic shielding even at subcritical current densities, originating from flux creep and thermally activated vortex motion. 
These results demonstrate that flux creep can substantially influence the nonlinear response of superconductors even under very weak AC drives~\cite{Basistha_2025}.
Despite these advances, prior studies have been restricted to collinear AC and DC magnetic fields, leaving the directional dependence of nonlinear magnetic screening under noncollinear or orthogonal field geometries largely unexplored.

Furthermore, in superconducting (SC) nanoelements, the geometry strongly affects vortex nucleation and pinning, for example, through the Bean–Livingston barrier. It also influences the interaction between vortices and screening currents. This interplay suggests that hysteretic properties and their anisotropy with respect to the orientation of the driving microwave field can be enhanced, giving rise to nontrivial and anisotropic nonlinear responses,
particularly in SHG in nanoscale SC elements. However, this aspect has not yet been explored.

To move beyond these limitations, we investigate the nonlinear microwave response of a type-II nanoscale SC prism, where geometrical confinement plays a significant role in the nonlinear dynamics and in the anisotropic relations between the dynamic magnetization and the driving field.
Our study explores the anisotropic response and analyzes the efficiency of SHG as a function of both the static bias field  and orientation of the driving field. 

The structure of the paper is as follows: 
Section~\ref{sec:model} introduces the physical model and the numerical approach based on the time-dependent Ginzburg--Landau framework. 
Section~\ref{sec:results} presents the simulation results and provides a detailed discussion of the underlying mechanisms governing the nonlinear magnetic response. 
Finally, Section~\ref{sec:conc} summarizes the main findings and outlines possible directions for future research.

%%%%%%%%%%%%%%%%%%%%%%%%%%%%%%%%%%%%%%%%%%%%%%%%%%%%%%%%%
\section{Model and Numerical Method\label{sec:model}}
\begin{figure}[t!]
  \centering
  \includegraphics[page=1,width=.48\textwidth]{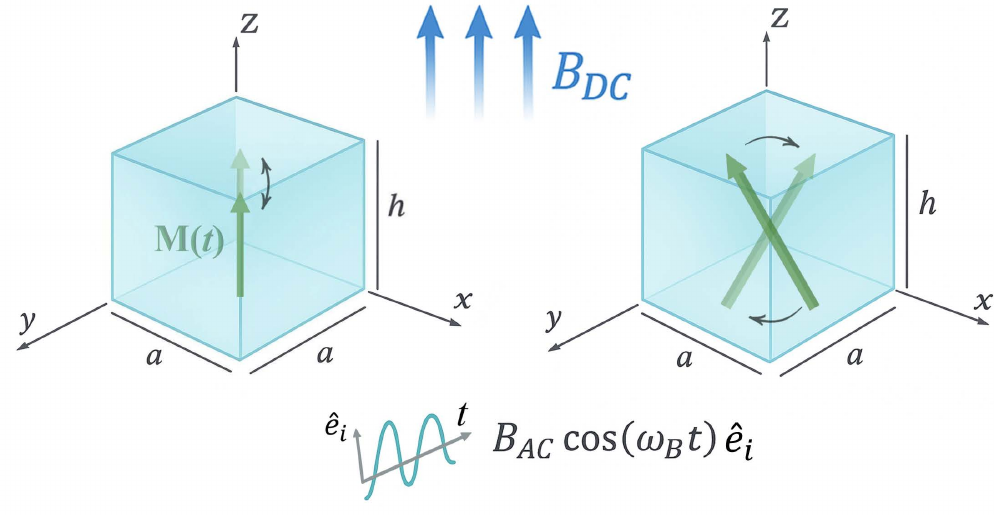}
  \put(-28,115){{\large{b}}}
  \put(-233,115){{\large{a}}}
\caption{
Schematic illustration of the SC sample and magnetic field configuration used in the simulations. The system is a type-II SC cube with side length $a=h=250$~nm, Ginzburg–Landau parameter $\kappa = 3$, and London penetration depth $\lambda=60$~nm. A uniform DC magnetic field is applied along the $z$-axis. 
To probe the nonlinear magnetic response, an AC magnetic field $ B_\text{AC}(t) = \Theta(t - t_0) \, B_\text{AC} \cos\left[\omega_{\text{B}} (t - t_0)   \right]$ is introduced, after the time $t_0$, once the system has reached a steady state under the influence of the DC bias field. The AC field is uniform in space and oscillates either in a) $z-$direction ($\hat{e}_i=\hat{e}_z$) or in b) $y-$direction ($\hat{e}_i=\hat{e}_y$).
}
\label{fig:structure}
\end{figure}

We consider a SC prism with dimensions $a \times a \times h$ along the $x$-, $y$-, and $z$-axes, respectively, as schematically shown in \textbf{Figure~\ref{fig:structure}}. As the main structure, we use a cube with $a=h=250$~nm, for which we perform most of the studies. The superconductor is characterized by a Ginzburg–Landau parameter $\kappa=\lambda/\xi=3$, where $\xi$ is the coherence length, a London penetration depth $\lambda=60$~nm, and a dimensionless electrical conductivity $\sigma=1$, expressed in units of $1/(\mu_0 D \kappa^2)$, where $D=10^5$~m$^2$/s is the diffusion constant.
The dynamic and static behavior of the SC system is modeled using the time dependent Ginzburg-Landau (TDGL) equations~\cite{Oripov2020, GROPP1996254, RevModPhys.74.99}. Originally derived for gapless superconductors~\cite{Gorkov, schmid}, the TDGL equations remain applicable to realistic superconducting materials 
in regimes where strong electron--phonon scattering smears the density-of-states singularities near the gap edge~\cite{lee_2023}. Today, TDGL is a standard approach for simulating superconductors, offering a practical balance between simplicity and accuracy~\cite{Horn_2023}.

Although most TDGL formulations employ fully dimensionless units, here we adopt a mixed scheme: the order parameter and conductivity are scaled dimensionlessly, whereas time, space, magnetic fields, and the vector potential remain in physical units~\cite{D4NH00618F, PhysRevE.101.033306, Alstrøm2011}. This choice preserves real-valued quantities directly related to the magnetic response of the system.
We solve the TDGL equations for a three-dimensional  type-II superconductor:
\begin{align}
    \frac{\xi ^2}{D}\frac{\partial\psi}{\partial t}
    &=-\frac{\lambda^2}{\kappa^2}(i \nabla+\frac{q}{\hbar}\mathbf{A})^{2} \psi+\psi-\lvert\psi\rvert^{2}\psi,\label{eq:psi2}
\\ 
    \sigma \frac{\xi ^2}{D} \frac{\partial\mathbf{A}}{\partial t}
    &=\frac{\hbar}{2iq}(\psi^{*}\nabla \psi-\psi\nabla\psi^{*})-\lvert{\psi}\rvert^{2}\mathbf{A}  
   -\lambda^2 \nabla\times(\nabla\times\mathbf{A}).
    \label{eq:vecpot2}
\end{align}
Here, $\psi$ is the order parameter and $\mathbf{A}$ is the magnetic vector potential.
Furthermore, $q = 2e$ is the Cooper pair charge, and $\hbar$ is the reduced Planck constant. 
The vector potential $\mathbf{A}$ represents the total magnetic field.
The $\nabla \times \mathbf{A}$ includes both the applied field, $\mathbf{B}_a$, and the screening field generated by superconducting currents.
Equation~\ref{eq:psi2},\ref{eq:vecpot2} were implemented in the mathematical module of \textsc{COMSOL Multiphysics}\textsuperscript{\tiny\textregistered}~\cite{zimmerman2006multiphysics, comsol2009, gulian2020shortcut} and solved using the Finite Element Method. The information about boundary conditions and further implementation details are provided in Supporting Information, section~S1.

The SC sample is subjected to a uniform static magnetic field $\mathbf{B}_{\text{DC}}$ along the $z$-axis and a uniform time-dependent field $\mathbf{B}_{\text{AC}}(t)$ applied along either the $y$- or $z$-axis:
\begin{align}\label{eq:Bfield}
\mathbf{B}_a(t) =
B_\text{DC} \, \hat{e}_z + B_\text{AC}(t) \, \hat{e}_i, 
\, i \in \{y, z\}, 
\end{align}
where $\hat{e}_i$ is the unit vector indicating the direction of the AC field. 
The AC component is turned on at time $t = t_0$, where $t_0 \gg 0$ ensures that the system has relaxed to a steady state under the DC field, reaching the equilibrium magnetization. The time dependence of the AC field is given by:
\begin{align}\label{eq:BfieldAC}
 B_\text{AC}(t) = \Theta(t - t_0) \, B_\text{AC} \cos\left[\omega_{\text{B}} (t - t_0)   \right] , 
\end{align}
where $\Theta(t - t_0)$ is the Heaviside step function, $B_{\text{AC}}$ is the amplitude and $\omega_{\text{B}}$ the angular frequency.
This setup ensures that the dynamic response to the AC excitation is isolated from transient relaxation effects associated with the initial application of the DC field.
The simulations are run for a sufficiently long time $t_1 \gg t_0$, allowing the system to reach a steady state and any transients induced by the AC field fully decay.

To evaluate the behavior of the SC system under applied magnetic fields, we analyze two key physical quantities: the time-dependent magnetization and the spatial distribution of vortices.

\paragraph{Magnetization:}  

We characterize the screening properties of the nanostructure by its diamagnetic response, using spatially averaged magnetization, which is quantified by  averaged magnetization over the volume of the superconductor~\cite{Berdiyorov_2009, nino_2019, gonzalez_2015, Aguirre_2017, hasnat_2020}:
\begin{equation}
\langle \mathbf{M} (\mathbf{r}, t) \rangle = \frac{1}{ V_{\text{SC}}} \int_{V_{\text{SC}}} \bigl[\nabla \times \mathbf{A} (\mathbf{r}, t) - \mathbf{B}_a (\mathbf{r}, t)\bigr] \, d^3\mathbf{r},
\label{eq:magnetisation}
\end{equation}
where $V_{\text{SC}}$ denotes the volume of the superconductor.

The SC sample is subjected to  $B_{\text{AC}}(t)$, and after the time $t_1$, the system reaches a stationary regime in which the magnetization oscillates around a new equilibrium configuration $\tilde{\mathbf{M}}_{\text{DC}}(\mathbf{r})$, which differs from the initial equilibrium magnetization $\mathbf{M}_{\text{DC}}$ due to the AC-induced shift. 
Accordingly, the volume-averaged magnetization can be expressed as:
\begin{align}
\langle \mathbf{M}(\mathbf{r}, t)\rangle &=
\begin{cases}
\langle \mathbf{M}_{\text{DC}}(\mathbf{r}) \rangle, & \text{for } t = t_0, \\[5pt]
\langle \tilde{\mathbf{M}}_{\text{DC}}(\mathbf{r}) \rangle + \langle \tilde{\mathbf{M}}_{\text{AC}}(\mathbf{r}, t) \rangle, & \text{for } t > t_1,
\end{cases}\label{eq:tam}
\end{align}
where the time-independent component, referred to as the effective DC magnetization, is defined as:
\begin{align}
\langle \tilde{\mathbf{M}}_{\text{DC}}(\mathbf{r}) \rangle &= \sum_{i \in \{x, y, z\}} \langle \tilde{{M}}_{i, \text{DC}}(\mathbf{r}) \rangle\, \hat{e}_i, \label{eq:tams}
\end{align}
while the dynamic component is given by:
\begin{align}
\langle \tilde{\mathbf{M}}_{\text{AC}}(\mathbf{r}, t) \rangle &=  \sum_{i \in \{x, y, z\}}\sum_{n} \tilde{{M}}_{i, n}\;
e^{[i n\omega_{B} (t - t_1) +\varphi_{i, n}]} \, \hat{e}_i,
\label{eq:tamd}
\end{align}
The effective DC magnetization is given by the zeroth Fourier component,
$\tilde{M}_{i, \mathrm{DC}} = \tilde{M}_{i,0}$, which corresponds to the
cycle-averaged value of 
$\langle \tilde{M}_{i,\mathrm{AC}}(\mathbf r,t)\rangle$.
This term represents the magnetization associated with the DC bias field,
including the static shift induced by the oscillating drive.
$\varphi_{i, n}$ denotes the phase shift of the $n^{\text{th}}$-order harmonic component of the magnetization with respect to the driving AC field, along the $i^{\text{th}}$ direction.
Each Cartesian component of the dynamic part was expressed as a Fourier series, where the complex amplitudes of the successive harmonics $\tilde{M}_{i, n}$ 
were calculated by the general formula:
\begin{align}\label{eq:harm}
&\tilde{M}_{i, n} = \\ & 
\left|
\frac{1}{2\pi} \int_0^{2\pi} 
\langle \tilde{M}_{i,\text{AC}}(\mathbf{r}, t) \rangle \, e^{-i n \omega_B (t-t_1) -\varphi_{i, n}} \, d(\omega_B (t-t_1))  \right|.\nonumber
\end{align}
The symbol $\langle \tilde{M}_{i,\text{AC}}(\mathbf{r}, t) \rangle$ denotes the volume-averaged dynamic component along the $i^{\rm th}$ Cartesian direction.
\paragraph{Vortex identification:} 
Vortex cores and regions of suppressed superconductivity are identified by analyzing the spatial distribution of the order parameter $|\psi|^2$. 
Because the transition between superconducting and 
normal-phases is gradual rather than sharply defined, a threshold value of $|\psi|^2 = 0.3$ is arbitrarily chosen to delineate vortex regions within the three-dimensional SC structure.

%%%%%%%%%%%%%%%%%%%%%%%%%%%%%%%%%%%%%%%%%%%%%%%%%%%%%%%%%

\section{\label{sec:results}Results and discussion}
\subsection{\label{sec:observations_DC}Response to the DC magnetic field
}
\begin{figure}[htp!]
  \centering
  \includegraphics[page=1,width=.48\textwidth]{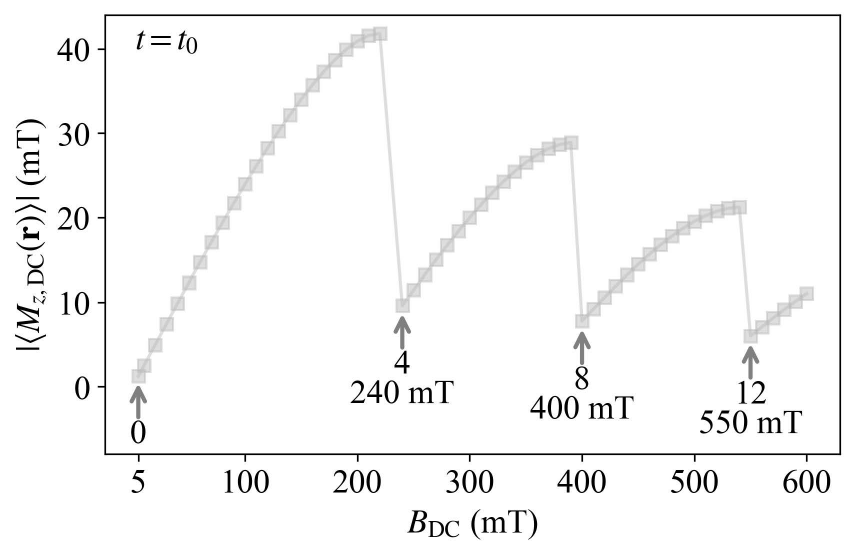}
\caption{
Volume-averaged $z$-component of DC magnetization,  $\langle M_{z,\,\text{DC}} \rangle$, the response of a SC cube as a function of the static and uniform magnetic field $B_{\text{DC}}\parallel\hat{z}$. Sharp discontinuities in curve, marked by gray arrows, indicate the nucleation of four vortices, at each specific field values: 240, 400 and 550~mT. 
}
\label{fig:M-Bstatic-basic}
\end{figure}
In homogeneous and isotropic superconductors, 
 the static magnetic response $\mathbf{M}_{\rm DC}$ follows the direction of the applied field $B_{\rm DC}$.
It is also supported by our simulations shown in \textbf{Figure~\ref{fig:M-Bstatic-basic}}, which presents the dependence of the averaged DC magnetization $\langle \mathbf{M}_{\text{DC}}(\mathbf{r}) \rangle=\langle M_{z, \text{DC}}(\mathbf{r}) \rangle \hat{e}_z$ on $B_{\text{DC}}$  in SC prism.
The transverse components of the averaged magnetization, 
i.e., $\langle M_{x, \text{DC}}(\mathbf{r}) \rangle$ and $\langle M_{y, \text{DC}}(\mathbf{r}) \rangle$, remain numerically negligible.

As $B_{\text{DC}}$ increases, the magnetization initially grows but exhibits a sequence of abrupt drops~\cite{PhysRevB.77.144509, PhysRevLett.81.2783, PhysRevLett.103.267002}, as indicated by gray arrows at 240, 400, and 550~mT. As the applied field approaches these critical values, the magnetization tends toward saturation. However, the maximal values of $\langle M_{z, \text{DC}}(\mathbf{r}) \rangle$ just before each drop systematically decrease from approximately 40~mT (at $B_\text{DC} = 230$~mT) to about 20~mT (at $B_\text{DC} = 540$~mT).
This nonlinear relationship between the static bias $B_{\text{DC}}$ and the static response $\langle M_{z,\text{DC}}(\mathbf{r}) \rangle$ reflects the presence of a critical Meissner current at which Abrikosov vortices nucleate in groups of four within the considered setup. The Meissner current saturates before reaching this critical value and then drops rapidly once vortices nucleate \cite{Tinkham1996, Bean_1964, Shmidt_1974}.
These nonlinearities become particularly relevant when  the static bias field $B_{\text{DC}}$ is supplemented by a dynamic field $B_{\text{AC}}$, as will be analyzed  in the following subsections.

%%%%%%%%%%%%%%%%%%%%%%%%%%%%%%%%%%%%%%%%%%%%%%%%%%%
\subsection{\label{sec:observations_AC} Mechanism of the SHG in combined DC and AC magnetic fields}
To illustrate how the SC cube responds to an AC drive, applied either parallel or transverse to the static field, we first examine a representative case, allowing us to include both the effects of vortex dynamics and the influence of Meissner currents on the AC response. We consider the DC field, $B_{\text{DC}} = 370$~mT $\parallel\hat{z}$,  when the system is in the mixed state  with four Abrikosov vortices (see Figure~\ref{fig:M-Bstatic-basic}).

\begin{figure*}[htp!]
  \centering
  {
      $\mathbf{B}_{\rm{DC}}~\parallel~\hat{z}, \;  \mathbf{B}_{\rm{AC}}~\parallel~\hat{z}$
      \hspace{6.2cm}
      $\mathbf{B}_{\rm{DC}}~\parallel~\hat{z}, \;  \mathbf{B}_{\rm{AC}}~\parallel~\hat{y}$}
  \includegraphics[page=1,width=.98\textwidth]{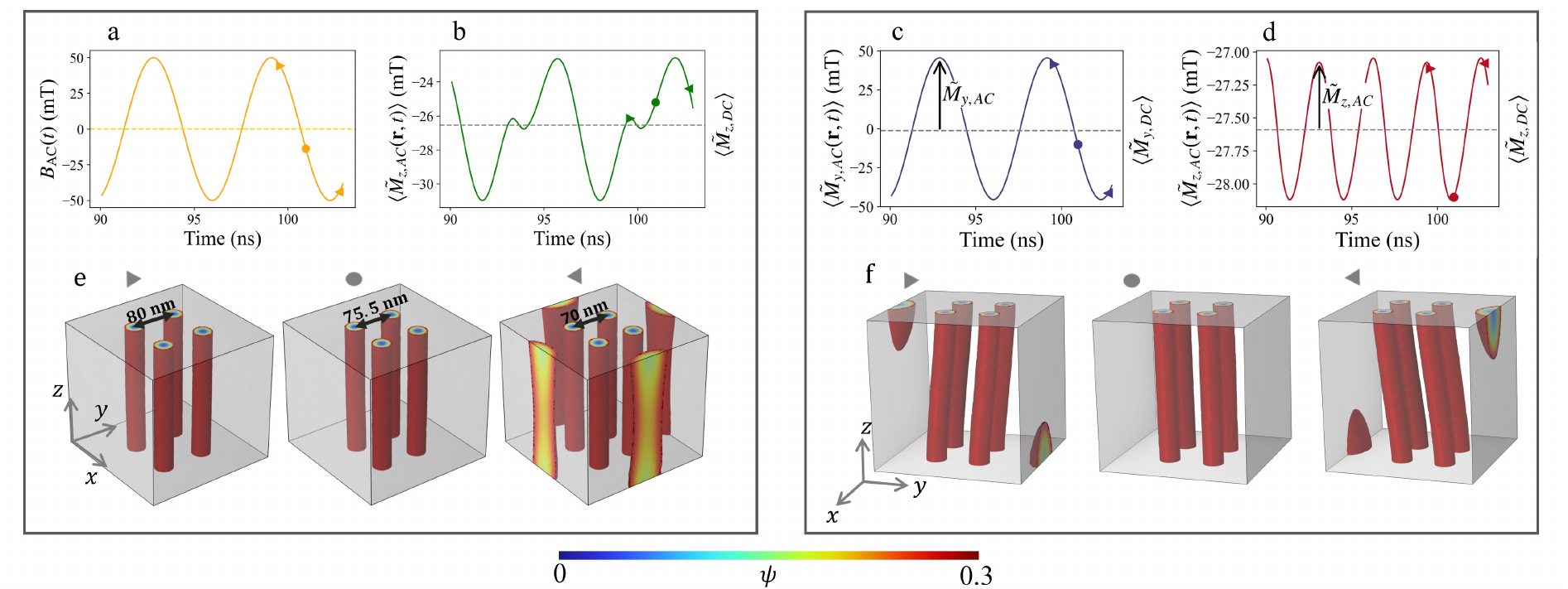}
\caption{
Time-resolved magnetization dynamics and vortex configurations under longitudinal or transverse AC magnetic field $B_{\mathrm{AC}}(t)$. 
Left panel: $\mathbf{B}_{\mathrm{AC}}\parallel\hat{z}$, a) applied AC field $B_{\mathrm{AC}}(t)$ and b) averaged magnetization in the $z$-direction, $\langle \tilde{M}_{z,\mathrm{AC}}(\mathbf{r},t) \rangle$.  
Right panel: $\mathbf{B}_{\mathrm{AC}}\parallel\hat{y}$, averaged magnetization in c) the $y$-direction, $\langle \tilde{M}_{y,\mathrm{AC}}(\mathbf{r},t) \rangle$, and d) the $z$-direction, $\langle \tilde{M}_{z,\mathrm{AC}}(\mathbf{r},t) \rangle$. The horizontal, gray dashed line in b), c) and d) indicates the effective DC magnetization, $\langle \tilde{\mathbf{M}}_{\text{DC}}(\mathbf{r}) \rangle$.
Symbols (circle, left- and right-pointing triangles) mark the AC field phases at which the vortex configurations are visualized, indicating the temporal evolution within an oscillation cycle.  
Snapshots of 3D vortex structures under e) longitudinal excitation show periodic compression and expansion of the inter-vortex spacing, while f) transverse excitation displays swinging-like oscillations of vortex lines.  
Simulation parameters: $B_{\mathrm{DC}} = 370$~mT ($\hat{z}$), $B_{\mathrm{AC}} = 50$~mT, $\omega = 1$~rad/ns, $\kappa = 3$, $\lambda = 60$~nm, $a=h=250$~nm.
 }
\label{3D: B(t) excitation}
\end{figure*}

%%%%%%%%%%%%%%%%%%%%%%%%%%%%%%%%%%%%%%
\begin{figure*}[htbp]
   \centering
     {
      $\mathbf{B}_{\rm{DC}}~\parallel~\hat{z}, \;  \mathbf{B}_{\rm{AC}}~\parallel~\hat{z}$
      \hspace{6.2cm}
      $\mathbf{B}_{\rm{DC}}~\parallel~\hat{z}, \;  \mathbf{B}_{\rm{AC}}~\parallel~\hat{y}$}
   \begin{tabular}{c }
       \includegraphics[page=1, width=.98\textwidth]{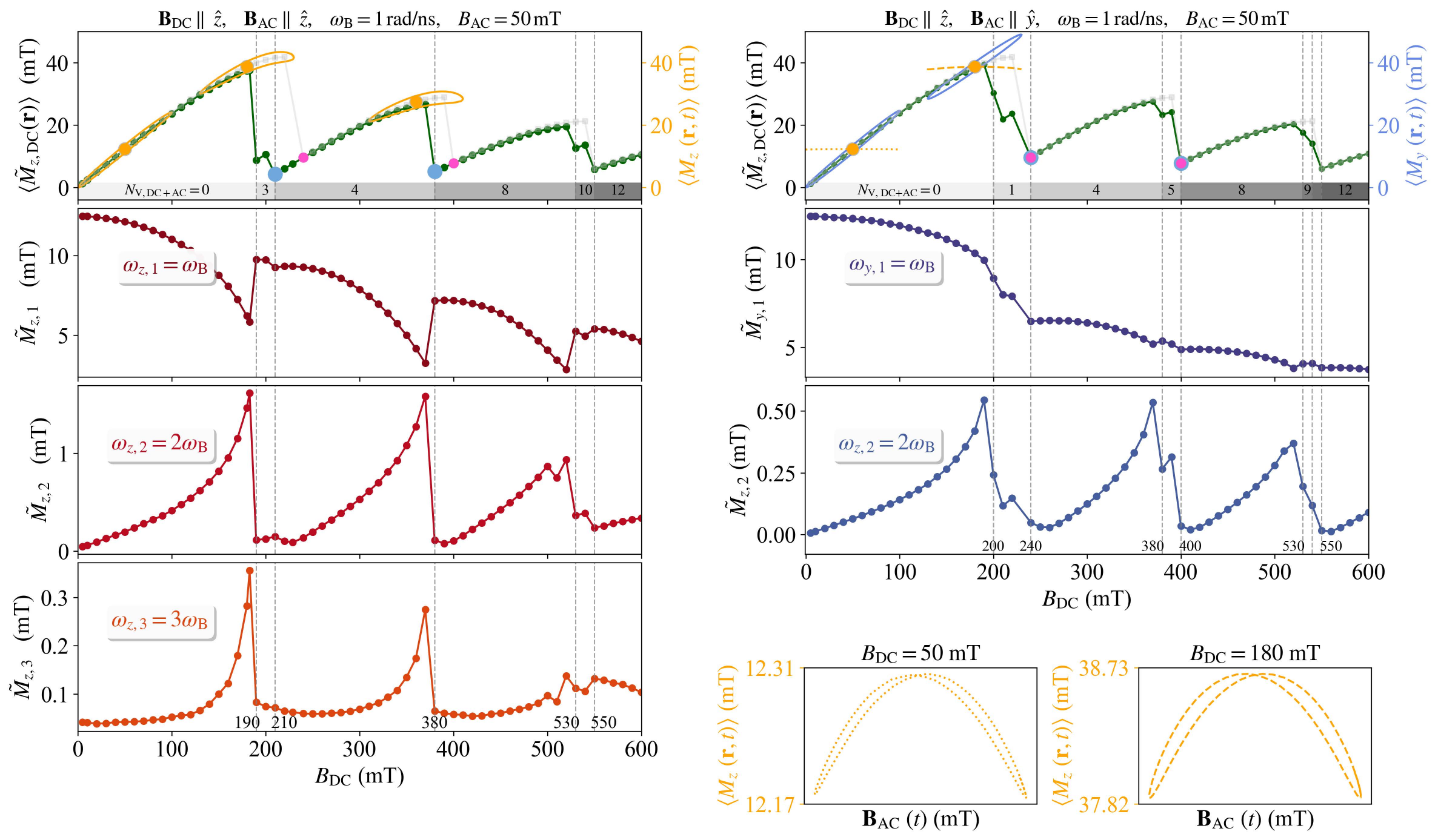}
       \put(-305,289){{{a}}}
       \put(-305,220){{{b}}}
       \put(-305,158){{c}}
       \put(-305,91){{{d}}}
      \put(-36,289){{{e}}}
      \put(-36,220){{{f}}}
      \put(-36,158){{{g}}}
      \put(-36,53){{{i}}}
      \put(-157,53){{{h}}}
   \end{tabular}
\caption{
Magnetic response of a SC cube under combined static and oscillating magnetic fields. The applied field is given by $\mathbf{B}(t) = B_{\rm DC} \, \hat{z} + B_{\rm AC}(t) \, \hat{e}_i$, with $i\!=\!y$ (right column) and $i\!=\!z$ (left column). The AC field has fixed amplitude $B_{\rm AC} = 50\,\mathrm{mT}$ and frequency $\omega_{\rm B} = 1\,\mathrm{rad/ns}$. a,e) Absolute value of the  $z$-component of the DC magnetization before the AC field is applied, $|\langle M_{z, \text{DC}} (\mathbf{r})\rangle|$ at $t\!=\!{t_0}$ (light gray line), and absolute value of the  $z$-component of the effective static magnetization after the AC field is applied, $|\langle \tilde{M}_{z, \text{DC}} (\mathbf{r})\rangle|$  at $t={t_1}$ (green line), plotted as a function of $B_{\rm DC}$. 
The light–blue circles indicate the DC field at which four vortices nucleate when the AC drive is present, while the light–pink circles denote the vortex-entry threshold under DC fields alone. 
b–d,f–g) Amplitude $\tilde{M}_{i, n}$ of the $n^{\mathrm{th}}$ harmonic of the oscillating magnetization component along direction $i$. Shaded regions indicate the vortex count $N_{V, {\rm DC+AC}}$. 
The hysteretic loops originated from the irreversible motion of the screening currents when the AC field is applied, are highlighted in panel a for the $z$–component of the magnetization (orange curves), and in panel e both for the  $y$–component (blue curves) and the $z$–component (orange curves). 
Panels h,i present magnified views of the $M_{z}$–$B_{\rm AC}$ hysteresis extracted from panel e, for representative bias fields $B_{\rm DC}=50~\mathrm{mT}$ and $B_{\rm DC}=180~\mathrm{mT}$, respectively. 
Material parameters: $\kappa = 3$, $\lambda = 60\,\mathrm{nm}$; cube side length $a = 250\,\mathrm{nm}$. 
}
\label{fig:delta_Mz-Bstatic}
\end{figure*}
%%%%%%%%%%%%%%%%%%%%%%%%%%%%%%%%%%%%%%%%%%%%%%%%%

When an AC field is applied along the $z$-axis (the left panel of \textbf{Figure~\ref{3D: B(t) excitation}}), the vortices exhibit a predominantly coherent, breathing motion: periodic contraction and expansion without altering their shape. Indentations with vortex seeds are formed at the centers of the side walls along the $z$-axis at the time when the AC field reaches its maximal positive value, as shown in Figure~\ref{3D: B(t) excitation}e.  This behavior maintains the symmetry of the square in the $x\text{-}y$ plane for both the distribution of Meissner currents and the alignment of the vortices.
Therefore, for a driving field applied along the $z$-direction, the longitudinal magnetization oscillates predominantly at the driving frequency while developing a nonsinusoidal waveform, indicating the presence of higher-order harmonics. As shown in Figure~\ref{3D: B(t) excitation}b, this component exhibits a clear phase delay relative to the driving field. The combination of waveform distortion and phase shift provides direct evidence of nonlinear longitudinal coupling.

Under transverse excitation, with the AC field applied along the $y$-axis, Figure~\ref{3D: B(t) excitation}f, the vortex cores in the middle of the cube remain fixed, while the vortex lines swing within the $y\text{-}z$ plane. 
Furthermore, normal-phase indentations grow and shrink in oscillatory way, at the midpoints of the top and bottom edges parallel to the $x$-axis.
This swinging dynamic of vortices and indentations breaks the inversion symmetry, 
distorts the circulation of current around the $z$-axis  and redistributes magnetic flux. As a consequence, the magnetization component along $y$, Figure~\ref{3D: B(t) excitation}c, oscillates at the first harmonic with the amplitude comparable to the $B_\text{AC}$ field, i.e., $\approx 50$~mT, Figure~\ref{3D: B(t) excitation}a, the 
$x$-component of the magnetization remains negligible ($<0.02$~mT), while the magnetization along $z$ exhibits frequency 
doubling, a clear signature of SHG, Figure~\ref{3D: B(t) excitation}d. This occurs despite the absence of direct driving along the $z$-direction, indicating intrinsic nonlinear anisotropic coupling driven by  broken symmetry.
In contrast to the collinear case, 
the transverse configuration produces oscillations in both $M_y$ and $M_z$ with negligible phase delay.

The emergence of first- and higher-order harmonics in the magnetization dynamics of the SC prism under the influence of a microwave magnetic field can be attributed to two mechanisms: vortex collective oscillations in the mixed state and screening currents associated with the Meissner background. Even in the specific example illustrated here, these two contributions are evident via vortex oscillations and normal-phase indentations.
In the following sections, we systematically analyze the magnetization dynamics as a function of both the amplitude of the applied AC and DC fields.

We consider a superconducting nanoelement (cube) with dimensions of a few London penetration depths, capable of hosting only a few vortices. Such a design exhibits strong nonlinearity, with the second harmonic dominating along the static-field direction under perpendicular driving. Although the absolute amplitude of the dynamic magnetization decreases when the superconductor size is only a few penetration depths, Supporting Information, section~S2 shows that,  for the cubic geometry, 
we already reach about $85\%$ of the  second-harmonic amplitude obtained in the limit $h \gg a$.

%%%%%%%%%%%%%%%%%%%%%%%%%%%%%%%%%%%%%%%%%%%%%%%%%%%
\subsection{Tuning SHG by adjusting the DC field\label{sec:SHG_DC}}
\begin{figure*}[htbp]
   \centering
     {\hspace{0.5cm}
      $\mathbf{B}_{\rm{DC}}~\parallel~\hat{z}, \;  \mathbf{B}_{\rm{AC}}~\parallel~\hat{z}$
      \hspace{1.3cm}}
    {
    \hspace{4.5cm}
      $\mathbf{B}_{\rm{DC}}~\parallel~\hat{y}, \;  \mathbf{B}_{\rm{AC}}~\parallel~\hat{z}$
      } 
   \begin{tabular}{c }
       \includegraphics[page=1, width=.98\textwidth]{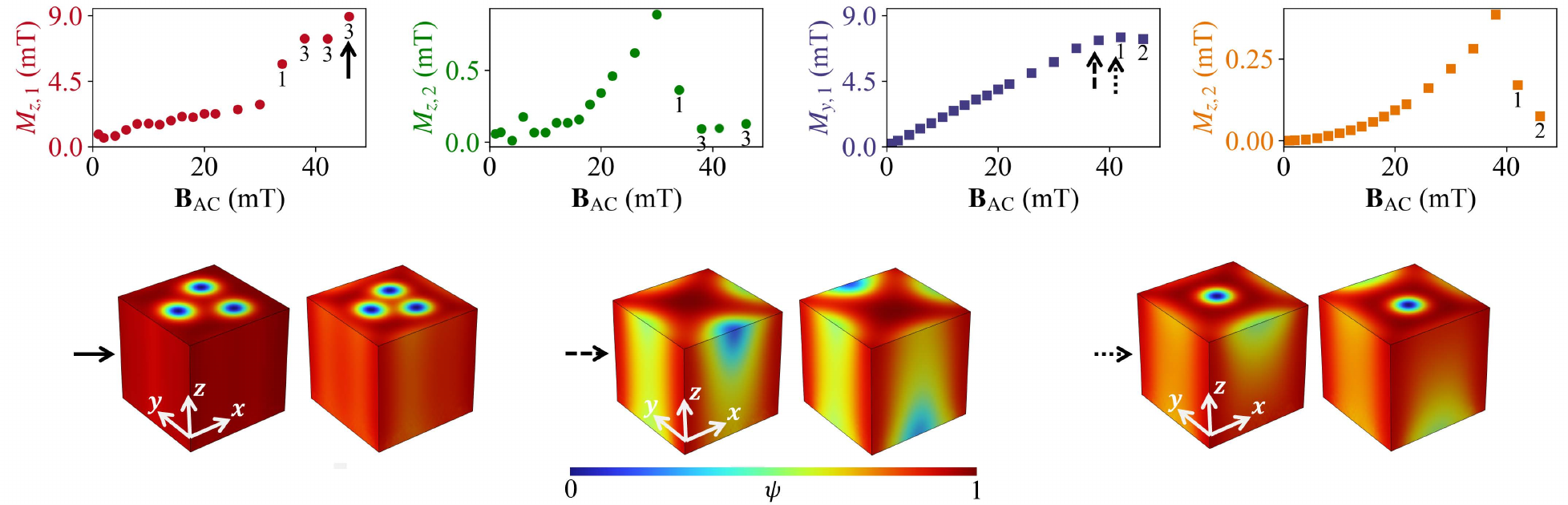}
       \put(-494,158){{{a}}}
       \put(-360,158){{{b}}}
       \put(-226,158){{c}}
       \put(-90,158){{{d}}}
      \put(-436,83){{{e}}}
      \put(-269,83){{{f}}}
      \put(-90,83){{{g}}}  
   \end{tabular}
\caption{
a–d) Dependence of the first- and second-harmonic magnetization amplitudes on the AC field strength, $B_{\mathrm{AC}}$, for fixed bias field, $B_{\mathrm{DC}} = 200$~mT.  
Two left plots correspond to longitudinal excitation ($\mathbf{B}_{\mathrm{AC}}\parallel\hat{z}$), while two right plots correspond to transverse excitation ($\mathbf{B}_{\mathrm{AC}}\parallel\hat{y}$).  
Panels a,c  show the first-harmonic amplitudes $M_{z,1}$ and $M_{y,1}$, respectively, while b,d display the second-harmonic components $M_{z,2}$.  
The numerical labels indicate the number of vortices observed in the steady-state configuration for each $(B_{\mathrm{DC}}, B_{\mathrm{AC}})$ pair.  
Bottom panels e-g show representative quasi-steady-state order-parameter distributions corresponding to the three drive conditions marked in the upper panels by solid, dashed, and dotted arrows, respectively.
Simulation parameters: $\omega_B = 1$~rad/ns, $\kappa = 3$, $\lambda = 60$~nm, $a = h = 250$~nm.
}\label{fig:B(t)}
\end{figure*}
\textbf{Figure~\ref{fig:delta_Mz-Bstatic}} summarizes the magnetic response of the SC cube subjected to an adjustable static magnetic field, $B_{\rm DC}$, applied along the $z$-axis, and to an AC magnetic field applied either longitudinally ($\mathbf{B}_{\rm AC}\parallel\hat{z}$, left column) or transversely ($\mathbf{B}_{\rm AC}\parallel\hat{y}$, right column). In both cases, $B_{\rm DC}$ is varied from 0 to 600~mT, while the AC field has a fixed amplitude of $B_{\rm AC} = 50$~mT and a frequency $\omega_\text{B} = 1$~rad/ns. All values presented in this figure are measured at $t = t_1 \gg t_0$, i.e., after the AC field is applied and the system has reached steady-state oscillations.

In Figure~\ref{fig:delta_Mz-Bstatic}a,e, the green lines show the absolute value of the 
effective static magnetization along the $z$-axis, $|\langle \tilde{M}_{z,\text{DC}}(\mathbf{r})\rangle|$, as a function of the bias magnetic field $B_{\text{DC}}$. 
As expected, the  dependencies of $|\langle \tilde{M}_{z,\text{DC}}
(\bf{r})\rangle|$ on $B_\text{DC}$ 
for both orientations of $\mathbf{B}_{\rm AC}$ are similar to those shown in Figure~\ref{fig:M-Bstatic-basic}, $|\langle {M}_{z,\text{DC}}
(\bf{r})\rangle|$, which are also placed in the background of Figure~\ref{fig:delta_Mz-Bstatic}a,e as light gray lines. 
However, adding an AC magnetic field decreases the critical
bias magnetic field at which successive sets of vortices start to nucleate. The first vortices nucleate at 
 $B_{\rm DC} = 190\,\mathrm{mT}$ for the $\mathbf{B}_{\rm AC}\parallel\hat{z}$ configuration and at $B_{\rm DC} = 200\,\mathrm{mT}$ for $\mathbf{B}_{\rm AC}\parallel\hat{y}$,
 while in the absence of $B_{\rm{AC}}$, it starts at 240~mT.
Nevertheless, the nucleation fields for 4-, and 8-vortex states are
reduced from $240\,\text{mT}$, and $400\,\text{mT}$ ($\mathbf{B}_{\rm AC} =0$) to approximately $210\,\text{mT}$, and $380\,\text{mT}$ for the  $\mathbf{B}_{\rm AC}\parallel\hat{z}$, and remain almost unchanged for $\mathbf{B}_{\rm AC}\parallel\hat{y}$, 
as quantified by the light blue and pink markers represented the 4$n$-vortex–entry thresholds with and without AC excitation, respectively (Figure~\ref{fig:delta_Mz-Bstatic}a,e).
 
The observed downshift of the nucleation field in the $\mathbf{B}_{\rm AC}\parallel\hat{z}$ configuration stems from the fact that, at a specific point in the AC field cycle, both components of the magnetic field sum up and surpass the critical value for vortex nucleation. Since vortex nucleation and annihilation are strongly hysteretic processes, the nucleated vortices remain in the system as the AC field decreases during its cycle. 
In the $\mathbf{B}_{\rm AC}\parallel\hat{y}$ configuration, the overlap of the DC and AC currents is weak (as discussed below), and the nucleation fields of the 4$n$-vortex remain essentially unchanged (Figure~\ref{fig:delta_Mz-Bstatic}e).
However, the nucleation of vortices with nonstandard numbers such as 1, 3, 5, etc., which occur at much smaller bias fields than 4$n$-vortex states, can be considered as metastable states induced by the $\mathbf{B}_{\rm AC}$ field. These states are very sensitive to the specific conditions of SC surfaces, such as the discretization mesh in the simulation or any defects in a real sample. 

Figure~\ref{fig:delta_Mz-Bstatic}b,f present the first harmonics of the SC response to the AC field in the $z$ and $y$-directions, i.e., the amplitudes $\tilde{M}_{z, 1}$ and $\tilde{M}_{y, 1}$, respectively (see~Equation~\ref{eq:harm}). The general trend of both  $\tilde{M}_{z,1}$, for $\mathbf{B}_{\mathrm{AC}}\parallel\hat{z}$, and $\tilde{M}_{y,1}$, for $\mathbf{B}_{\mathrm{AC}}\parallel\hat{y}$, is similar: they decrease with increasing $B_{\rm DC}$.
This decay is a result of the decreasing volume fraction of the SC phase as the number of vortices increases. 
Consequently, the dynamic diamagnetic response of the SC phase 
generally diminishes with increasing $B_{\rm DC}$.
However, with the nucleation of vortices at successive $B_{\rm DC}$ fields, we observe an increase in the amplitude of the first harmonic, and it is observed only when the AC driving field is applied along the $z$-direction, Figure~\ref{fig:delta_Mz-Bstatic}b. 
This is because, for $\mathbf{B}_{\rm AC}$ oscillating along the $z$-direction, the static and dynamic Meissner currents strongly overlap: both circulate or oscillate predominantly on the lateral faces of the SC cube, around the $z$-axis. 
Thus, as the DC current component saturates with increasing $B_{\rm DC}$ and approaching the critical value for vortex nucleation, the amplitude of the AC current, added on top of the static one, is also reduced. After vortices are nucleated and the static component of the current drops abruptly, the dynamic component can recover, as it is no longer impeded by saturation. This reduction and recovery of the AC response before and after vortex entry is clearly visible in Figure~\ref{fig:delta_Mz-Bstatic}b. By contrast, a similar trend is barely observable in Figure~\ref{fig:delta_Mz-Bstatic}f for an AC field applied along $y$, where the AC and DC currents do not overlap strongly and the DC saturation only weakly affects the AC contribution.

The main nonlinear dynamic response of the SC cube to the AC magnetic field is shown in Figure~\ref{fig:delta_Mz-Bstatic}c,g for $\mathbf{B}_{\mathrm{AC}}\parallel\hat{z}$ and $\mathbf{B}_{\mathrm{AC}}\parallel\hat{y}$, respectively. It shows the amplitude of the $\tilde{M}_{z,2}$, which represents the $z$-component of the magnetization oscillating at the double frequency of the driving AC field, as a function of $B_\text{DC}$. Interestingly, SHG is observed for both orientations of the AC field at the $z$-component of the magnetization, and the dependencies are qualitatively very similar, only with different amplitudes, reaching 1.6~mT and 0.6~mT in Figure~\ref{fig:delta_Mz-Bstatic}c,g, respectively.  
The amplitude $\tilde{M}_{z,2}$ exhibits exponential growth, interspersed by abrupt discontinuous drops at the bias fields corresponding to vortex nucleation.

The emergence and evolution of higher harmonics in the magnetic response of the SC prism are fundamentally due to the nonlinear terms in the TDGL equation~\ref{eq:psi2},\ref{eq:vecpot2}, such as $|\psi|^2\psi$ and $|\psi|^2\mathbf{A}$. 
A clear confirmation is obtained by comparing TDGL solutions with the London-limit dynamics (see Supporting Information section~S3). In the London approximation, 
where the order-parameter amplitude is kept rigid and the equations are linear (Equation~S6 in Supporting Information), 
the magnetization remains purely sinusoidal and no higher harmonics appear. 
To provide an intuitive picture of the nonlinear magnetic response observed in Figure~\ref{fig:delta_Mz-Bstatic}c,g, we will consider the supercurrents,
since magnetization is linearly related to the current.
 As the DC component of the supercurrent begins to saturate, entering the nonlinear regime, the nonlinearity of the AC-current oscillations about this large DC offset also becomes enhanced. This nonlinearity is manifested by the non-elliptical shape of the minor loops in the dependence of magnetization on the total field in Figure~\ref{fig:delta_Mz-Bstatic}a,e, and in h-i.
In Figure~\ref{fig:delta_Mz-Bstatic}a, the minor loops, 
representing the response to the $z$-oriented AC field, are well 
formed, indicating the dominance of the first harmonic. However, their asymmetric shapes point at a  significant contribution from both even and odd harmonics. 
In contrast, the dynamic response shown with minor hysteresis loops in   Figure~\ref{fig:delta_Mz-Bstatic}e is more peculiar. 
The yellow minor loops seem to be flat,
however, they reveal a symmetric omega-like shape when zoomed in, as shown in Figure~\ref{3D: B(t) excitation}h-i. Their amplitude range of $\langle \tilde{M}_{z} (\mathbf{r},t)\rangle$ increases from 0.14~mT at $\mathbf{B}_{\mathrm{DC}}=50$~mT to 0.90~mT at $\mathbf{B}_{\mathrm{DC}}=180$~mT. This shape, typical of Lissajous curves with a 1:2 frequency ratio \cite{Lawrence_72}, demonstrates nonlinearity in the anisotropic AC response.
Consequently, the first and other odd harmonics of the AC response are suppressed, and the most significant contribution is given by the second harmonic of the $z$-component of AC magnetization  induced by the AC field applied along the $y-$direction. 
It is worth noting that the nonlinearity of the AC response along the driving field, i.e., $y$-axis, is  suppressed (the blue minor loops remain nearly elliptical), 
which is due to the partial spatial separation of DC and AC currents in this system (see the discussion in Figure~\ref{3D: B(t) excitation}). This makes the anisotropic SHG a dominant nonlinear response in the $\mathbf{B}_{\rm AC}\parallel\hat{y}$ configuration.

We showed in Figure~\ref{fig:delta_Mz-Bstatic}c,g that the SHG is most pronounced for values of $B_{\rm DC}$ close to the vortex-nucleation thresholds.
As discussed earlier, in this regime the Meissner current becomes large and eventually saturates. However, this saturation is additionally accompanied by the formation of normal-phase indentations at the centers of the side faces of the SC cube. 
 The path along which the Meissner currents circulate is then dynamically perturbed because the Meissner current bypasses these indentations, whose sizes change with the applied AC field.
 These changes are nonlinear due to the nonlinear dynamics of the order parameter $\psi$. Consequently, the nonlinearity of the current dynamics, and the associated magnetization dynamics, increases as the DC field approaches the vortex-nucleation threshold, with the DC field biasing the AC excitation into this strongly nonlinear regime. 

%%%%%%%%%%%%%%%%%%%%%%%%%%%%%%%%%%%%%%%%%%%%%%%%
\subsection{Impact of the amplitude of applied AC field on~SHG}

In subsection~\ref{sec:SHG_DC}, we discussed the system response to variations in the value of the static component, $B_{\mathrm{DC}}\hat{\mathbf{z}}$, of the external field while keeping the amplitude of its dynamic component (the driving field), $B_{\mathrm{AC}}$, fixed. We now examine how the dynamic magnetization changes when the static field is set to $B_{\mathrm{DC}} = 200~\mathrm{mT}$, close to the threshold for the nucleation of the first vortices, and the amplitude $B_{\mathrm{AC}}$ is varied.

\textbf{Figure~\ref{fig:B(t)}} summarizes the numerical results for longitudinal ($\mathbf{B}_{\mathrm{AC}}\parallel\hat{z}$, see Figure~\ref{fig:B(t)}a,b,e) and transverse ($\mathbf{B}_{\mathrm{AC}}\parallel\hat{y}$, see Figure~\ref{fig:B(t)}c,d,f,g) driving. Panels Figure~\ref{fig:B(t)}a--d present the responses, i.e. the amplitudes of the first and second harmonics of the dynamic magnetization, as functions of $B_{\mathrm{AC}}$. The first harmonics (Figure~\ref{fig:B(t)}a,c)  are shown along the direction of the driving field. The second harmonics (Figure~\ref{fig:B(t)}b,d) are taken along the direction in which they are nonzero: either parallel to the driving field $\mathbf{B}_{\mathrm{AC}}$ (Figure~\ref{fig:B(t)}b) or perpendicular to $\mathbf{B}_{\mathrm{AC}}$ (Figure~\ref{fig:B(t)}d).

In the regime of small driving amplitudes, i.e., up to vortex nucleation fields, the first harmonic of dynamic magnetization is proportional to $\mathbf{B}_{\rm AC}$, but the slope of $\tilde{M}_{1,z}$ is smaller than that of $\tilde{M}_{1,y}$.
This is consistent with the discussion already provided in the subsection \ref{sec:SHG_DC}. Namely, for $\mathbf{B}_{\rm AC}\parallel\hat{\mathbf{z}}$, the static and dynamic Meissner currents flow through the same region; therefore, saturation of the DC current blocks the growth of the AC current amplitude. In contrast, for $\mathbf{B}_{\rm AC}\parallel\hat{\mathbf{y}}$ the DC and AC current paths only partially overlap, and the increase of $\tilde{M}_{1,y}$ is not inhibited. For the same reasons, the vortex nucleation occurs at different AC amplitudes in the two field geometries, 34~mT for $\mathbf{B}_{\rm AC}\parallel\hat{z}$ and 42~mT for $\mathbf{B}_{\rm AC}\parallel\hat{y}$).

The second harmonic amplitude $\tilde{M}_{2,z}$ is a nonlinear function of the driving field amplitude  $B_{\rm AC}$. This dependence seems to be  parabolic for $\mathbf{B}_{\rm AC}\parallel\hat{y}$, however, for $\mathbf{B}_{\rm AC}\parallel\hat{z}$, it is less regular. For parallel and perpendicular driving, the second harmonic $\tilde{M}_{2,z}$  reaches the highest amplitudes of 0.99~mT and 0.40~mT, respectively.  Such values are observed for the amplitudes of driving fields approaching the threshold for vortex nucleation. 

Nevertheless, whenever the vortices are nucleated, despite the very large amplitude of $B_{\rm AC}$, the vortex dynamics, whether exhibiting breathing behavior (Figure~\ref{fig:B(t)}e) or swinging behavior (Figure~\ref{fig:B(t)}f,g), does not generate a strong second harmonic. These results are consistent with previous observations that the nonlinear response is diminished once the Meissner current is reduced. This also supports our conclusion that the SHG is maximized when the Meissner currents are saturated and the normal-phase indentations oscillate due to the AC field.

\section{Conclusions\label{sec:conc}}

We investigated the effect of an oscillating magnetic field on the nonlinear magnetic response of a nanoscale superconducting cube. The diamagnetic screening response exhibits second-harmonic generation  in the dynamic magnetization. The nonlinearity is enhanced when the static bias field is just below one of the critical values for Abrikosov-vortex nucleation. This enhancement is attributed to the saturation of the Meissner current and its perturbation by the nonlinear dynamics of the normal-phase indentations.

Importantly, the resulting second-harmonic generation cannot be explained by conventional vortex-based mechanisms, such as vortex depinning or thermally activated vortex motion. Instead, it reflects a distinct nonlinear regime that emerges from  symmetry breaking imposed by the noncollinear field configuration.

A particularly interesting feature is the dynamic component of the magnetization parallel to the bias field when the AC field is applied perpendicularly. In this anisotropic nonlinear response, only even harmonics appear; therefore, the leading spectral component is the second harmonic, which is not overshadowed by the frequency of the driving field.
Overall, our findings reveal a nonlinear mechanism for anisotropic SHG in confined superconductors and suggest new strategies for controlling nonlinear electrodynamic effects in nanoscale SC devices, with potential applications in quantum sensing, cryogenic electronics, and tunable SC elements.

%\vspace{1 cm}
\section*{Author contributions}
The project was conceived by J.W.K. and M.K., who also supervised the research. 
S.M. designed and implemented the simulation methodology, performed the numerical analysis. 
A.G. contributed additional supervision and scientific input. 
All authors participated in interpreting the results and revising the final manuscript.

%\vspace{1 cm}
\section*{Conflicts of interest}
There are no conflicts to declare.

%\vspace{1 cm}
\section*{Acknowledgements}
The authors would like to thank K.~Szulc for his valuable assistance with the numerical simulations.
The work was supported by the grants of the National Science Center – Poland, No. UMO-2021/43/I/ST3/00550 (SM and JWK) and UMO-2020/39/I/ST3/02413 (MK). 

%\vspace{1 cm}
\section*{Data Availability}
Data supporting this study are openly available from the repository at webpage:
\\
{https://doi.org/10.5281/zenodo.18145617}.

%\vspace{1 cm}

%\medskip
\section*{Supporting Information} \par Supporting Information is available from the Wiley Online Library and contains the following sections:\break S1~--~Details of theoretical model, 
S2 -- Role of the shape of the SC nanoelement
S3 -- London approximation of the TDGL equation
S4 -- TDGL Model --  ac response in the Meissner state.
\vspace{0.5cm}

% Acknowledgements
%\medskip
%\textbf{Acknowledgements} \par %delete if not applicable))
%Please insert your acknowledgements here

% References
\medskip
\clearpage

% Use the following code if you wish to generate your bibliography with BibTeX;
% replace the string "MSP-template" below with the name(s) of
% the BibTeX data base(s) you want to use.
% The resulting bibliography-output (the content of the .bbl file)
% must be pasted back into this file before submission.
% Please also include your BibTeX data base file(s) in your submission
% so that we can re-run BibTeX if necessary.
%
%\bibliographystyle{MSP}
%\bibliography{biblio}

% \textbf{References}\\

%merlin.mbs apsrev4-1.bst 2010-07-25 4.21a (PWD, AO, DPC) hacked
%Control: key (0)
%Control: author (8) initials jnrlst
%Control: editor formatted (1) identically to author
%Control: production of article title (-1) disabled
%Control: page (0) single
%Control: year (1) truncated
%Control: production of eprint (0) enabled
%

\clearpage


\section*{Supplementary material}

# Supporting Information:
# Anisotropic second harmonic generation in superconducting nanostructures

Sara Memarzadeh,[1, *] Maciej Krawczyk,[1] Armen Gulian,[2] and Jarosław W. Kłos[1]

[1]*Institute of Spintronics and Quantum Information,*
*Faculty of Physics and Astronomy, Adam Mickiewicz University Poznań,*
*Uniwersytetu Poznańskiego 2, 61-614 Poznań, Poland*
[2]*Laboratory of Advanced Quantum Materials and Devices,*
*Institute for Quantum Studies, Chapman University,*
*One University Drive, Orange, California, 92866, United States of America*

## S1: DETAILS OF THEORETICAL MODEL

The superconducting (SC) state near the critical temperature is conveniently described by the phenomenological time-dependent Ginzburg–Landau (TDGL) formalism, which extends the static GL theory to dynamical regimes. In this framework, the functional derivatives of the GL free-energy functional with respect to the order parameter $\psi^*(\mathbf{r},t)$ and the vector potential $\mathbf{A}(\mathbf{r},t)$ provide the driving forces that relax the system toward lower free-energy configurations. Accordingly, the TDGL equation governing the temporal evolution of the order parameter is [1]:

$$\frac{\hbar^2}{2m^*D}\frac{\partial \psi(\mathbf{r},t)}{\partial t} = -\left[\alpha\psi(\mathbf{r},t) + \beta|\psi(\mathbf{r},t)|^2\psi(\mathbf{r},t) \right.$$
$$\left. + \frac{1}{2m^*}\left(\frac{\hbar}{i}\nabla - q\mathbf{A}(\mathbf{r},t)\right)^2\psi(\mathbf{r},t)\right], \quad \text{(S1)}$$

where all the parameters are defined in the main text. Similarly, the magnetic vector potential $\mathbf{A}(\mathbf{r},t)$ evolves according to Maxwell's equations, modified to include both the superconducting current and the applied magnetic field. Denoting $\sigma$ as the conductivity, we can describe the time evolution of $\mathbf{A}(\mathbf{r},t)$ via Equation S2 which is valid for uniform applied field $\mathbf{B}_a$:

$$\sigma\frac{\partial \mathbf{A}}{\partial t}(\mathbf{r},t) = \frac{q\hbar}{2m^*i}\left(\psi^*(\mathbf{r},t)\nabla\psi(\mathbf{r},t) - \psi(\mathbf{r},t)\nabla\psi^*(\mathbf{r},t)\right)$$
$$- \frac{q^2}{m^*}|\psi(\mathbf{r},t)|^2\mathbf{A}(\mathbf{r},t) - \frac{1}{\mu_0}\nabla\times\nabla\times\mathbf{A}(\mathbf{r},t). \quad \text{(S2)}$$

It is important to note that the analysis presented in the Equation S1 and S2 is based on a specific choice of gauge with $\Phi = 0$, thereby eliminating the time-dependent scalar potential. For the simulations, we also made the complex order parameter $\psi$ and the electrical conductivity $\sigma$ dimensionless [2]:

$$\psi \to \psi_0\psi, \quad |\psi_0|^2 = \frac{|\alpha|}{\beta}, \quad \sigma \to \frac{1}{\mu_0 D\kappa^2}\sigma, \quad \text{(S3)}$$

---

[*] sara.memarzadeh@amu.edu.pl

We use the real-dimensional vector potential $\mathbf{A}$, and the spatial coordinates $(x,y,z)$ to ensure proper coupling to a microwave magnetic field. With these transformations, the TDGL equations for the order parameter $\psi$ and the vector potential $\mathbf{A}$ become:

$$\frac{\xi^2}{D}\frac{\partial \psi}{\partial t} = -\frac{\lambda^2}{\kappa^2}(i\nabla + \frac{q}{\hbar}\mathbf{A})^2\psi + \psi - |\psi|^2\psi, \quad \text{(S4)}$$

$$\sigma\frac{\xi^2}{D}\frac{\partial \mathbf{A}}{\partial t} = \frac{\hbar}{2iq}(\psi^*\nabla\psi - \psi\nabla\psi^*) - |\psi|^2\mathbf{A}$$
$$- \lambda^2\nabla\times\nabla\times\mathbf{A}. \quad \text{(S5)}$$

We employ COMSOL Multiphysics®[3] to implement the time-dependent Ginzburg-Landau (TDGL) equations using the *Coefficient Form PDE* interface, which provides a flexible framework for defining custom equations. The TDGL equations, Equation S5 and S4, are solved with a time-dependent solver, assuming negligibly small electrical conductivity outside the superconducting domain.

At the superconducting boundary, the conditions $\nabla\psi\cdot\mathbf{n} = 0$ and $\mathbf{A}\cdot\mathbf{n} = 0$ (where $\mathbf{n}$ is the unit vector normal to the boundary) are imposed to prevent any current from flowing across it [4]. Additionally, at distances far from the superconductor, larger than ten times the SC prism size, the magnetic field is assumed to approximate the applied field $\mathbf{B}_a$, i.e., $\nabla\times\mathbf{A} = \mathbf{B}_a$.

## S2: ROLE OF THE SHAPE OF THE SC NANOELEMENT

To investigate how geometric confinement influences the generation of higher harmonics, we analyze the dependence of the magnetization response on the sample height $h$, i.e. as $h$ increases from 250 nm to 2 $\mu$m, we transit from the SC prism to the long rod of square cross-section. For these systems, we consider fixed amplitude of transverse excitation ($\mathbf{B}_{\text{AC}} = 50$ mT $\parallel \hat{y}$) and fixed DC field ($\mathbf{B}_{\text{DC}} = 150$ mT $\parallel \hat{z}$). **Figure** S1 shows the first-harmonic transverse component $M_{y,1}$ and the second-harmonic amplitude $M_{z,2}$ as a functions of $h$. Since these fields are below the vortex-entry threshold, the system remains in the Meissner state and the magnetic response is entirely determined by surface screening currents.

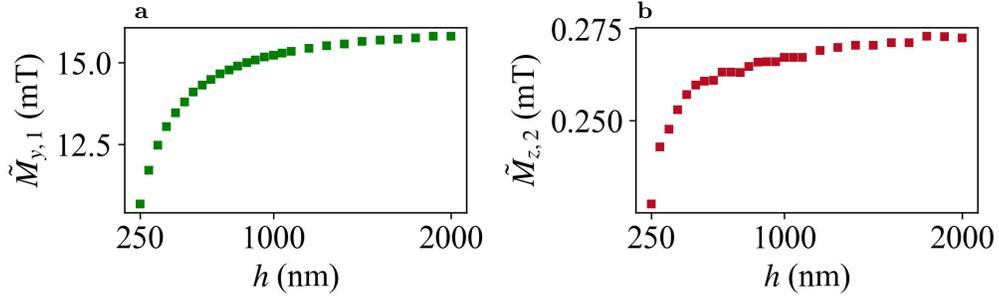

FIG. S1. a) First ($M_{y,1}$, green)- and b) second ($M_{z,2}$, red)-harmonic of the dynamic magnetization as a functions of the superconducting higth $h$, varied from 250 nm to 2 $\mu$m, under transverse AC driving ($\mathbf{B}_{\text{AC}} \parallel \hat{y}$) and a uniform DC field ($\mathbf{B}_{\text{DC}} \parallel \hat{z}$). The simulations were performed for $\mathbf{B}_{\text{DC}} = 150$ mT, $\mathbf{B}_{\text{AC}} = 50$ mT, and $\omega_B = 1$ rad/ns.

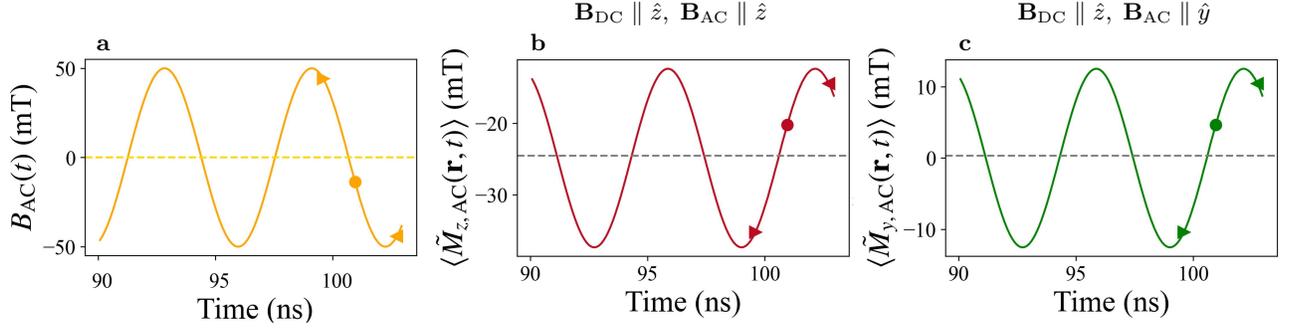

FIG. S2. Time-resolved response on the a) applied AC magnetic field. b,c) The induced magnetization components are obtained using the London approximation ($\psi = 0$), which simplifies the TDGL model $\psi = 1$. The static field is fixed at $B_{\text{DC}} = 100$ mT $\parallel \hat{z}$, while the AC excitation is applied b) longitudinally, $B_{\text{AC}} = 50$ mT $\parallel \hat{z}$, or c) transverse to the DC field, $B_{\text{AC}} = 50$ mT $\parallel \hat{y}$. The dashed lines indicate the corresponding effective static magnetization $\langle \tilde{M}_{i,\text{DC}}(\mathbf{r}, t) \rangle$. All other dynamic components of the magnetization are zero within the numerical accuracy.

Both the first-harmonic response ($M_{y,1}$) and the induced second harmonic ($M_{z,2}$) exhibit a monotonic increase with increasing higth, followed by saturation once $h$ becomes sufficiently large. This trend is not related to vortex dynamics, absent in this regime, but instead arises from how the AC field is screened inside the prism.

Because of the finite London penetration depth of the SC currents, the diamagnetism of superconductors is not perfect. Meissner currents do not flow strictly at the surface; therefore, the effectively screened volume is smaller than the geometric volume of the sample, which reduces the diamagnetic response This general mechanism of incomplete screening can used to explain saturation of dynamical responses, presented in Figure S1.

The $y$- and $z$-components of the dynamic magnetization are associated with oscillatory screening currents circulating around the $y$- and $z$-axes, respectively. We first consider the $z$-component, which in the case of transverse driving is dominated by the second harmonic, $M_{z,2}$, as shown in Figure S1b. The Meissner currents circulating around the $z$-axis are reduced along these trajectories which are situated near the bottom and top faces of the prism. Because this reduction is local, $M_{z,2}$ saturates as the prism height $h$ increases, and further increases in $h$ have only a weak effect on the signal.

The $y$-component of the dynamic magnetization is dominated by the first harmonic, $M_{y,1}$. The saturation of $M_{y,1}$ (see Figure S1a) can be understood as follows: as $h$ increases, the corresponding Meissner-current paths become longer, while the contribution from its sections close the top and bottom faces of SC prism becomes progressively less significant to the overall screening.

For a cube with $h = 250$ nm, we have a relatively large ratio $h/\lambda \approx 4.17$, and the first and second harmonics reach about 67% and 85% of their saturation values in the limit of large $h$. Therefore, the cubic geometry is a well-justified choice for SHG.

## S3: LONDON APPROXIMATION OF THE TDGL EQUATION

**Figure** S2 shows the magnetization response obtained within the London approximation. Assuming $\psi = 1$, for uniform applied field $\mathbf{B}_a$, equation governing the vector potential reduces to:

$$\sigma \frac{\xi^2}{D} \frac{\partial \mathbf{A}}{\partial t} = -\mathbf{A} - \lambda^2 \nabla \times \nabla \times \mathbf{A} \qquad (S6)$$

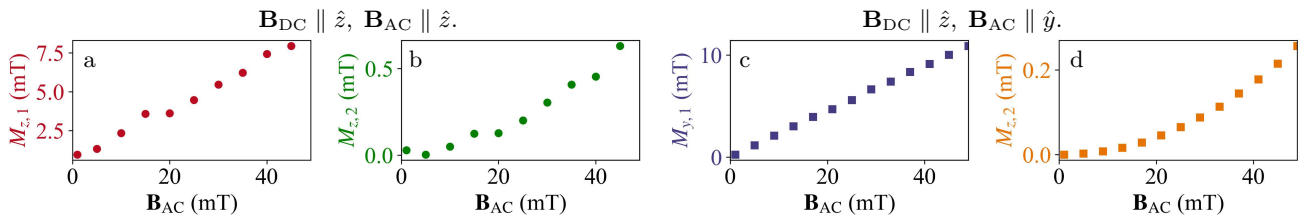

FIG. S3. a–d) Dependences of the first- and second-harmonic of dynamic magnetization on the AC field strength $B_{AC}$ for fixed bias field $B_{DC} = 150$ mT. Two left plots a,b) correspond to longitudinal excitation ($\mathbf{B}_{AC} \parallel \hat{z}$), while two right plot c,d) refer to transverse excitation ($\mathbf{B}_{AC} \parallel \hat{y}$). Panels a,c) show the first-harmonic amplitudes $M_{z,1}$ and $M_{y,1}$, respectively, while b,d) display the second-harmonic components $M_{z,2}$. Simulation parameters: $\omega_B = 1$ rad/ns, $\kappa = 3$, $\lambda = 60$ nm, $a = h = 250$ nm.

This relation resembles a damped wave equation and shows how the magnetic vector potential changes in space and time inside a superconductor.

Thus, within the London approximation, the intrinsic nonlinearities of the full TDGL formalism are eliminated. Consequently, the magnetization dynamics respond linearly to the applied AC field, and no higher harmonics appear in the spatially averaged magnetization components. This is fully consistent with wave forms shown in Figure S2, where all the induced magnetization components with non-negligible amplitudes follow sinusoidal oscillations at the drive frequency, in contrast to the TDGL simulations in the main text, where nonlinearities generate pronounced higher-order harmonic content. It is worth noting that within the London approximation, the system does not exhibit an anisotropic response, i.e. the dynamic magnetisation in the $z-$direction while the field is aligned with the $y-$direction. This is because, in the nonlinear model based on the TDGL equations, the first harmonic of the anisotropic response is not observed. Therefore, after reducing the TDGL formalism to London theory, the anisotropic response is lost completely.

## S4: TDGL MODEL – AC RESPONSE IN THE MEISSNER STATE

To complement the discussion of the results presented in Figure 5 in the main text, focused on the regime close to the vortex-nucleation threshold at $B_{DC} = 200$ mT, we now analyze the dynamic magnetic response of the superconducting cube for a smaller static field, $B_{DC} = 150$ mT. In this regime, the system remains in the Meissner state for all the driving amplitudes shown in **Figure** S3, and no vortices are nucleated throughout the entire AC cycle.

For both excitation geometries, longitudinal ($\mathbf{B}_{AC} \parallel \hat{z}$) and transverse ($\mathbf{B}_{AC} \parallel \hat{y}$), the first harmonic of the dynamic magnetization ($\tilde{M}_{z,1}$ or $\tilde{M}_{y,1}$) increases nearly linearly with $B_{AC}$. This behavior is characteristic of reversible Meissner screening, where the AC and DC shielding currents coexist without reaching the saturation regime that would otherwise suppress the linear response.

The second harmonic $\tilde{M}_{z,2}$ remains smaller than in the $B_{DC} = 200$ mT case discussed in the main text. This reduced SHG arises because the external magnetic field never approaches the values required to generate mature normal-core indentations that act as the vortex seeds. Although the system stays entirely in the Meissner state, a parabolic increase of $\tilde{M}_{z,2}$ with $B_{AC}$ persists. This originates from the intrinsic nonlinear terms of the TDGL equations, which produce a finite nonlinear response even when the applied field is far below the critical nucleation threshold.



\end{document}